\documentclass[]{IAC_style}

\usepackage{algorithm}
\usepackage{algpseudocode}
\usepackage{amsmath}

\usepackage{caption}  

\makeatletter
\renewcommand{\ALG@name}{}  
\makeatother

\begin{document}

\IACpaperyear{2024} 
\IACpapernumber{IAC-24-D5.4.2-x87772} 
\IAClocation{Milan, Italy} 
\IACdate{14-18 October 2024} 

\IACcopyrightA{}

\title{Encrypted Computation of Collision Probability for Secure Satellite Conjunction Analysis}

\IACauthor{Jihoon Suh$^{\orcidlink{0000-0000-0000-0000}}$}{1}{1}
\IACauthor{Michael Hibbard$^{\orcidlink{0000-0000-0000-0000}}$}{1}{0}
\IACauthor{Kaoru Teranishi$^{\orcidlink{0000-0000-0000-0000}}$}{1}{0}
\IACauthor{Takashi Tanaka$^{\orcidlink{0000-0000-0000-0000}}$}{1}{0}
\IACauthor{Moriba Jah$^{\orcidlink{0000-0000-0000-0000}}$}{1}{0}
\IACauthor{Maruthi Akella$^{\orcidlink{0000-0000-0000-0000}}$}{1}{0}

\IACauthoraffiliation{Department of Aerospace Engineering and Engineering Mechanics, The University of Texas at Austin \\ \normalfont{E-mail:\ \texttt{jihoonsuh@utexas.edu, mwhibbard@utexas.edu, teranishi@utexas.edu, ttanka@utexas.edu, moriba@utexas.edu, makella@mail.utexas.edu}}}

\abstract{
The computation of collision probability ($\mathcal{P}_c$) is crucial for space environmentalism and sustainability by providing decision-making knowledge that can prevent collisions between anthropogenic space objects. However, the accuracy and precision of $\mathcal{P}_c$ computations is often compromised by limitations in computational resources and data availability. While significant improvements have been made in the computational aspects, the rising concerns regarding the privacy of collaborative data sharing can be a major limiting factor in the future conjunction analysis and risk assessment, especially as the space environment grows increasingly privatized, competitive, and fraught with conflicting strategic interests. This paper argues that the importance of privacy measures in space situational awareness (SSA) is underappreciated, and regulatory and compliance measures currently in place are not sufficient by themselves, presenting a significant gap.

To address this gap, we introduce a novel encrypted architecture that leverages advanced cryptographic techniques, including homomorphic encryption (HE) and multi-party computation (MPC), to safeguard the privacy of entities computing space sustainability metrics, inter alia, $\mathcal{P}_c$. Our proposed protocol, Encrypted $\mathcal{P}_c$, integrates the Monte Carlo estimation algorithm with cryptographic solutions, enabling secure collision probability computation without exposing sensitive or proprietary information. This research advances secure conjunction analysis by developing a secure MPC protocol for $\mathcal{P}_c$ computation and highlights the need for innovative protocols to ensure a more secure and cooperative SSA landscape.

}

\maketitle
\thispagestyle{fancy} 



\section*{Acronyms/Abbreviations}
\noindent The following acronyms and abbreviations are used:
\begin{itemize}
    \item LEO: Low Earth Orbit
    \item HE: Homomorphic Encryption
    \item SSA: Space Situational Awareness
    \item USSPACECOM: The United States Space Command
    \item ASAT: Anti-Satellite
    \item DSS: Distributed Satellite Systems
    \item FHE: Fully Homomorphic Encryption
    \item MPC: Multi-party Computation
\end{itemize}

\section{Introduction}
The reliance on satellite and spacecraft technologies has become essential to modern society, impacting vital sectors and applications across various domains, such as navigation, banking, agriculture, strategic planning, and scientific research. \cite{pelton2015new, wertz2011space}. However, the rapid expansion in the use of space has led to an alarming increase in space debris orbiting the Earth, creating a congested orbital environment that significantly raises the risk of collisions \cite{kessler1978collision, liou2011active}. These collisions can result in the permanent loss of highly valued assets like communication satellites, weather satellites, navigation systems, Earth observation instruments, and scientific research equipment. The direct consequences include disruption of global communications, impaired weather forecasting, compromised navigation and timing systems, hindered agricultural planning and disaster response, and setbacks in scientific progress. The crash between an inactive Russian satellite, Cosmos 2251, and an active U.S. commercial satellite, Iridium 33, \cite{Jakhu2010} reminds us of the potential consequences of miscalculated collision risks. This collision resulted in more than 1,800 pieces of space debris exceeding 10 cm and larger as well as many thousands of smaller fragments. Much of this debris will remain in orbit until the end of this century, posing ongoing risks to other objects in Low Earth Orbit (LEO), \cite{johnson2009collision}, and the increasing space debris from such collisions accelerates the saturation of the finite orbital carrying capacity \cite{palmroth2021toward} that would eventually lead to an unsustainable environment for future space activities. Preserving the sustainable orbital environment for future generations is an important shared responsibility \cite{lawrence2022case}, and we aim to contribute to these broader initiatives by promoting international collaboration and ethical practices through technological innovation.

It is clear that prioritizing the prevention of such catastrophes, not only to protect valuable assets but also to curb the growth of space debris for a more sustainable orbital environment. The computation of collision probability ($\mathcal{P}_{c}$) is one of the most important tools in assessing and managing the collision risks \cite{nasa_ca} and has been recognized early on, leading to the developments of rigorous analysis and efficient computation methods \cite{alfriend1999probability, akella, patera2001general, 6224727}. Moreover, the aforementioned collision incident also urged national organizations such as the U.S. Space Command (USSPACECOM) to conduct a space situation awareness (SSA), and organizations in charge are recommended to share and exchange the information on the position of objects in orbit among program partners \cite{weeden2010iridium}. Such cooperative measures are more important today because the orbital environment is a shared environment occupied by diverse stakeholders, such as national space agencies, defense organizations, and commercial satellite operators. 

Findings in \cite{dolan2023satellite} demonstrated that sharing high-quality information among operators significantly improves coordination and reduces the risk of collisions. However, it also postulates that privacy concerns and adversarial relationships between satellite operators can lead to limited information sharing, leading to suboptimal coordination and increased collision risk. A notable example is the recent near-collision event between China's Tiangong space station and Starlink satellites in 2021, which highlighted how geopolitical tensions and the lack of transparency can exacerbate collision risks \cite{jones2021chinastarlink}. Policymakers have introduced detailed regulatory frameworks to address these data-sharing challenges, but relying solely on such legal measures is often insufficient. Even when relations are non-adversarial, many participants view their orbital information as proprietary or strategic. They perceive the need to share this data as an invasion of privacy or a serious security threat, particularly in light of anti-satellite (ASAT) and other counter-space capabilities demonstrated by major state actors in the past \cite{green2014space, pardini2009assessment, samsonspace}, and more recently by the developing Russian ASAT system \cite{kimball2024us}. 

While regulatory frameworks encourage more extensive data sharing, the technological solutions addressing the accompanying security and privacy concerns remain largely undeveloped. This paper proposes a novel secure protocol that enables the privacy-preserved computation of collision probability $\mathcal{P}_c$ through the combination of multiple cryptographic primitives, aiming to alleviate privacy and security concerns and foster an environment where cooperation among stakeholders becomes more natural and advantageous.

\subsection{Literature Review}
The first MPC protocol specifically developed for the secure satellite conjunction analysis was introduced by \cite{hemenway2014achieving}. In this work, the authors explored the feasibility of implementing the available MPC protocols within the honest-but-curious adversarial model \footnote{In the honest-but-curious (semi-honest) adversarial model, all parties adhere to the established protocols precisely (honest), but attempt to learn as much as possible (curious) about other parties' inputs from the data accessible to them. This model is often a starting point for developing a new protocol, and is progressively enhanced for more powerful adversaries.}, at the time by analyzing the circuit complexity of Alfano's method \cite{alfano2005numerical} for computing $\mathcal{P}_c$, and determined time estimates for a high precision implementation. The follow-up work \cite{hemenway2016high} by the same authors presented the concrete design tailored to compute $\mathcal{P}_c$ under the MPC framework, and extended its security analysis from the honest-but-curious threat model to the semi-malicious. Using the Sharemind \cite{bogdanov2008sharemind}, which is a framework for secure MPC with secret-sharing, \cite{kamm2015secure} also demonstrated the secure evaluation of the $2$-D integral expression for computing $\mathcal{P}_c$ between two satellites.

The space industry is revisiting Distributed Satellite Systems (DSS) due to their potential to enhance resilience, coverage, and scalability. Several technical gaps exist that hinder the full realization of DSS, and \cite{selva2017distributed} highlights the critical need for research in data integrity, encryption, and in-space network resilience to enable secure data sharing among satellite operators.

Secure dataspace approach \cite{hanke2024secure} builds on the aforementioned foundational works by providing an outline of how the integral form of $\mathcal{P}_c$ can be securely computed using the elementary functions provided in the MP-SPDZ library \cite{keller2020mp} while the inputs to those functions are securely processed by utilizing secret-sharing and HE; more importantly, authors of \cite{hanke2024secure} examines how these secure computation methods can be operationalized within a dataspace infrastructure, providing a logistical foundation for the practical implementations in the future.

In short, previous works share a common theme where the use of various cryptographic primitives, e.g., Secure MPC, SS, and HE facilitates the successful evaluation of the analytical integral computing the $\mathcal{P}_c$ for the same objective: privacy-preserving satellite conjunction analysis. While these earlier works have primarily focused on the direct evaluation of the integral expression for $\mathcal{P}_c$, our research introduces a novel approach by bringing the Monte Carlo simulation approach into the secure MPC design, which leads to a different solution and perspective on calculating $\mathcal{P}_c$ in a secure manner.

Our key contribution is the development of a novel secure satellite conjunction analysis protocol, Encrypted $\mathcal{P}_c$. This protocol, to the best of our knowledge, is the first of its kind to securely perform Monte Carlo simulations for computing $\mathcal{P}_c$ by leveraging the combination of HE and MPC framework, enhancing security and privacy while providing a robust alternative to existing integral-based approaches found in the literature. We propose a series of subprotocols that, together, achieve the confidential computation of $\mathcal{P}_c$ in the three-party computation environment that consists of two independent satellite operators, and a cloud server.

\subsection{Organization}
This paper is organized as follows. In section \ref{prelim}, we review the basic computation procedures for $\mathcal{P}_c$, and identify the potential privacy compromise within the secure conjunction analysis, which motivates our proposed method. Section \ref{method} presents the proposed Encrypted $\mathcal{P}_c$ protocol $\Pi$ in detail, starting with the protocol's architecture, threat model, and review of cryptographic primitives that constitute the main protocol. We conclude with Section \ref{discussion} which offers discussions, future work, and a summary of our study.

\section{Preliminaries}\label{prelim}
\subsection{Brief on Computation of $\mathcal{P}_c$} \label{Pc_review}
In this section, we briefly describe conventional methods\cite{nasa_ca} for computing $\mathcal{P}_c$ and examine how the procedure necessitates the sharing of potentially sensitive data that is not intended to be disclosed for security reasons, such as precise positions and uncertainties.

$\mathcal{P}_c$ of two satellites is a critical metric in assessing the risk of conjunctions, guiding decisions for collision avoidance maneuvers. The computation of \(\mathcal{P}_c\) hinges on accurately estimating the likelihood that the distance between two objects at their closest approach will be less than a specified threshold, usually the sum of their radii.

This computation procedure builds upon several key assumptions, justified by the brief duration of the encounter events. First, we assume that the nominal trajectories of the spacecraft can be represented by straight lines with constant velocities. Velocity uncertainties at the point of encounter are neglected. The position uncertainties, however, are modeled as constant and follow uncorrelated Gaussian distributions. Furthermore, the primary object is assumed to be significantly larger than the secondary object, allowing the latter to be treated as a point mass. Both spacecraft are also assumed to be spherical in shape.

To calculate \(\mathcal{P}_c\), we first need to establish the positions and velocities of the two spacecraft at the time of closest approach. The position and velocity vectors of the primary and secondary objects, $r_1, r_2, v_1$, and $v_2$, respectively, serve as the foundational inputs and are defined as:
\begin{equation}\label{pos}
r_1 = \begin{bmatrix}x_1 & y_1 & z_1\end{bmatrix}^{\top}, \; r_2 = \begin{bmatrix}x_2 & y_2 & z_2\end{bmatrix}^{\top}
\end{equation}
\begin{equation}
v_1 = \begin{bmatrix}v_{1}^{x} & v_{1}^{y} & v_{1}^{z}\end{bmatrix}^{\top}, \; v_2 = \begin{bmatrix}v_{2}^{x} & v_{2}^{y} & v_{2}^{z}\end{bmatrix}^{\top}
\end{equation}

The uncertainties in positions, represented by their covariance matrices \(\Sigma_1\) and \(\Sigma_2\), capture the imprecision in the objects' locations. These matrices are given by:
\begin{equation}
\Sigma_i = \mathbb{E}\left[(r_i - \bar{r}_i)(r_i - \bar{r}_i)^{\top}\right], \;\; i=1,2
\end{equation}

With the position and velocity vectors defined, the next step is to consider the physical dimensions of the spacecraft. The hard-body radius \(R\), which is the sum of the radii of the primary and secondary objects, quantifies the minimum distance at which a collision would occur:
\begin{equation}
R = R_1 + R_2
\end{equation}
Given these inputs, we proceed to compute the combined covariance matrix \(\Sigma_c\), which aggregates the positional uncertainties of both objects:
\begin{equation} \label{combined_cov}
\Sigma_c = \Sigma_1 + \Sigma_2
\end{equation}

To frame the encounter in a more convenient coordinate system, we define the relative position and velocity vectors, \(r_{rel}\) and \(v_{rel}\), respectively, and the specific angular momentum \(h\):
\begin{equation} \label{rel}
r_{rel} = r_1 - r_2, \quad v_{rel} = v_1 - v_2, \quad h = r_{rel} \times v_{rel}
\end{equation}

These vectors allow us to construct a reference transformation matrix \(Q\), which aligns the new coordinate system with the relative motion of the objects:
\begin{equation} \label{transform}
Q = \begin{bmatrix}X & Y & Z\end{bmatrix}^{\top},
\end{equation}
\noindent where
\[Y = \frac{v_{rel}}{\|v_{rel}\|}, \quad Z = \frac{h}{\|h\|}, \quad  X = Y \times Z.
\]

This transformation is applied to the combined covariance matrix, resulting in the transformed covariance matrix \(\Sigma_{c, \text{XYZ}}\):
\begin{equation}\label{cov_transform}
\Sigma_{c, \text{XYZ}} = Q \Sigma_c Q^{\top}
\end{equation}

We then project this transformed covariance onto the XZ-plane (the conjunction plane), yielding the 2D covariance matrix \(\Sigma_{c, \text{XZ}}\):
\begin{equation}\label{cov_proj}
\Sigma_{c, \text{XZ}} = \begin{bmatrix}1 & 0 & 0 \\ 0 & 0 & 1\end{bmatrix}\Sigma_{c, \text{XYZ}}\begin{bmatrix}1 & 0 & 0 \\ 0 & 0 & 1\end{bmatrix}^{\top}
\end{equation}

Combining \eqref{cov_transform} and \eqref{cov_proj}, we have a single transformation
\begin{equation}\label{cov_transform_and_project}
\Sigma_{c, \text{XZ}} = Q_{\text{XZ}}\Sigma_{c}Q_{{\text{XZ}}}^{\top},
\end{equation}
\noindent where \[Q_{XZ} = \begin{bmatrix}1 & 0 & 0 \\ 0 & 0 & 1\end{bmatrix} Q.\]

Next, we calculate the distance \(r_0\) between the two objects in the secondary object's reference frame, as well as the projected miss vector \(r_{\text{XZ}}\):
\begin{equation}
r_0 = \|r_{rel}\|, \quad r_{\text{XZ}} = \begin{bmatrix}1 & 0 & 0 \\ 0 & 0 & 1\end{bmatrix}r_{rel}
\end{equation}

Finally, the collision probability \(\mathcal{P}_c\) can be computed by integrating over the conjunction plane as follows:
\begin{equation}\label{2dintegral}
\mathcal{P}_{c} = \dfrac{1}{2\pi |\Sigma_{c, \text{XZ}}|^{\frac{1}{2}}}\int_{LB_x}^{UB_x}\int_{LB_z}^{UB_z} f(x, z) \, dz \, dx,
\end{equation}
\noindent where
\begin{align*}
LB_x &= r_0 - R, \: UB_x = r_0 + R, \\
LB_z(x) &= -\sqrt{R^2 - (x - r_0)^2}, \: UB_z(x) = \sqrt{R^2 - (x - r_0)^2} \\
f(x, z) &= \exp\left(-\frac{1}{2}r_{\text{XZ}}^{\top}\Sigma_{c, \text{XZ}}^{-1}r_{\text{XZ}}\right)
\end{align*}

\subsection{Monte Carlo Estimation of $\mathcal{P}_c$}
\begin{figure}[h] 
    \centering
    \includegraphics[width=0.48\textwidth]{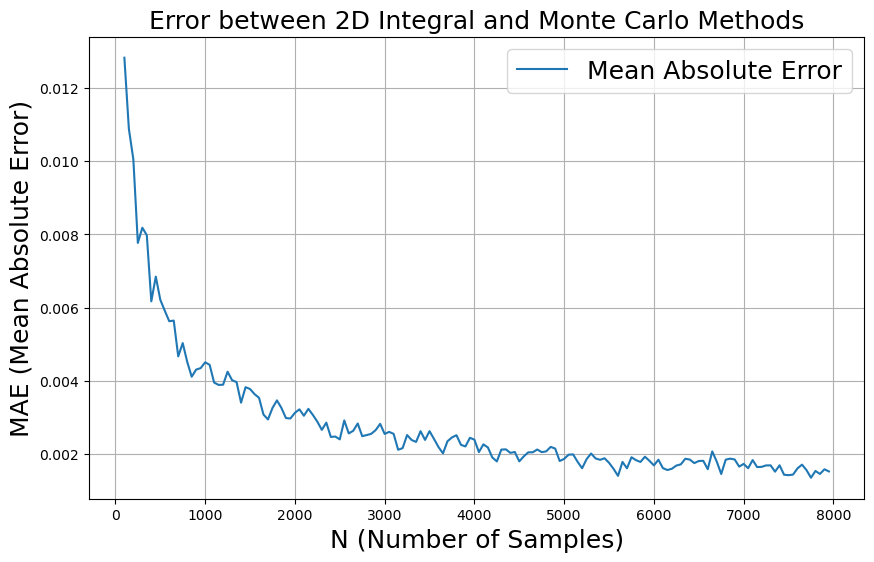}
    \caption{The mean absolute error (MAE) between the integral and the Monte Carlo approaches for computing $\mathcal{P}_c$.}
    \label{fig2-1:MC_accuracy}
\end{figure}

As an alternative to the 2D Integral approach presented in the preceding subsection, the Monte Carlo method offers a flexible and powerful way to estimate the probability of collision without the need to compute the analytical integral in \eqref{2dintegral}. This method leverages stochastic sampling to approximate the probability of collision by simulating the potential positions of the objects involved with its error decaying proportionate to the inverse square root of the number of samples, $N$, as seen in Figure~\ref{fig2-1:MC_accuracy}.

\subsubsection{Relative Position, Velocity, and Covariance}
The process begins by calculating the quantities such as the combined hard-body radius, combined positional covariance, and the relative position and velocity vectors as in \eqref{combined_cov}-\eqref{rel}. As in the 2D integral approach, a new coordinate frame is defined, centered at the combined covariance, and aligned with the relative position and velocity vectors. These transformations are identical to the ones described earlier in the 2D integral approach. Similarly, the combined covariance matrix $\Sigma_c$ is also transformed and then projected onto the conjunction plane to obtain the 2D covariance matrix \(\Sigma_{c, \text{XZ}}\). The process and corresponding equations are the same as in the 2D integral approach and thus can be referred to by Equations \eqref{cov_transform} and \eqref{cov_proj}.

\subsubsection{Monte Carlo Simulation}
\begin{figure}[h] 
    \centering
    \includegraphics[width=0.48\textwidth]{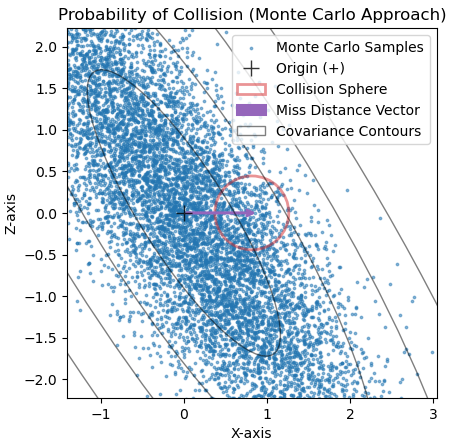}
    \caption{Monte Carlo Simulation to compute $\mathcal{P}_c$}
    \label{fig2-2: PcMC}
\end{figure}

The Monte Carlo method involves generating $N$ random samples from a bivariate normal distribution with the covariance matrix $\Sigma_{c, \text{XZ}}$: $$s^i \sim \mathcal{N}\left(0, \Sigma_{c, \text{XZ}}\right), \: i=1, \cdots, N.$$ These samples represent potential positions in the conjunction plane, as illustrated in Figure~\ref{fig2-2: PcMC}.

The number of samples that fall within the collision region, defined by a collision sphere centered at $(r_0, 0)$ with radius $r_0$, is counted, i.e., $n_c$. The probability of collision $\mathcal{P}_c$ can be estimated by dividing $n_c$ by the number of samples $N$:
\begin{equation}
    \mathcal{P}_c \approx \frac{n_c}{N}
\end{equation}

\subsection{Motivation for Secure Computation}
The accurate computation of $\mathcal{P}_c$ often requires collaboration between different stakeholders such as national space agencies, defense organizations, and commercial satellite operators. In many cases, these stakeholders must share sensitive data, including precise positional and uncertainty information (as seen in Section \ref{Pc_review}) to jointly assess the collision risk. However, the need for such precise data sharing is fraught with challenges as they can be strategically sensitive. Exposed orbital data can be exploited by adversaries for ASAT attacks. Commercial satellite operators are also not without risks as their proprietary operational data could reveal competitive intelligence, such as the operational status and the expected lifespan of their spacecraft. The complex nature of international relations in space operations means that trust is often limited, complicating efforts to collaborate on the basis of legal and regulatory frameworks only.

These concerns relate to the well-known privacy-utility dilemma \cite{li2009tradeoff} that extends beyond our focus on the encryption of $\mathcal{P}_c$, and is ubiquitous in numerous applications requiring privacy enhancement. In the context of our research, this dilemma manifests as follows: On one hand, accurate collision probability computations demand proactive and comprehensive data exchange; on the other, sharing such sensitive information introduces significant security and proprietary risks, clearly calling for secure and reliable computation protocols that can balance the privacy and utility. To address this challenge, we propose Encrypted $\mathcal{P}_c$ as a potential solution, leveraging HE and MPC frameworks to ensure data confidentiality during $\mathcal{P}_c$ computation. 

\section{Proposed Methodology: Encrypted $\mathcal{P}_c$} \label{method}
\subsection{Overview of the Proposed Approach}
While HE provides a powerful theoretical framework for ensuring the confidentiality of data and the computation itself, it also confines us to the elementary operations: addition and multiplication. Such computational restriction presents a new challenge when designing a privacy-preserving protocol. Traditional methods for computing $\mathcal{P}_c$ involve operations that do not naturally align with the arithmetic limitations of HE, necessitating a more creative approach. For this reason, we carefully re-engineer the existing Monte Carlo simulation approach to develop an encrypted MPC protocol that can compute $\mathcal{P}_c$ while keeping the satellite operator's data private.

\subsubsection{Multi-Party Architecture and Threat Model}
\begin{figure}[h]
    \hspace*{-0.15cm}
    \includegraphics[width=0.48\textwidth]{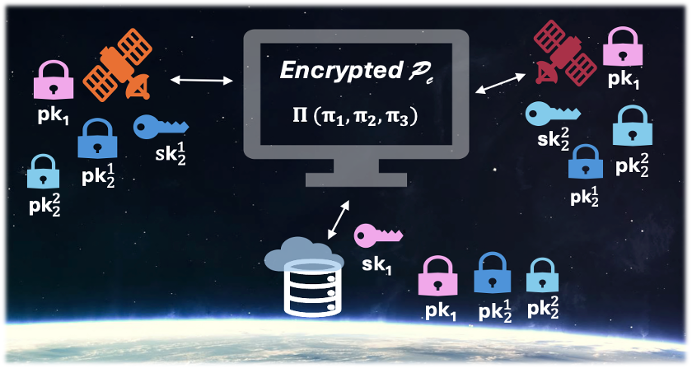}
    \caption{Architecture of Encrypted $\mathcal{P}_c$ protocol: a series of subprotocols $\pi_{i=1, \cdots, 3}$ make up the entire protocol $\Pi$, which is jointly run by three parties: orange (left) and maroon (right) satellite operators (Operator $1$ and $2$) each holding their secret keys $\mathsf{sk}_{2}^{1}$ and $\mathsf{sk}_{2}^{2}$, respectively, and the coordinator (cloud server) holding $\mathsf{sk}_1$ to assist the protocol. Public keys ($\mathsf{pk}_{1}$, $\mathsf{pk}_{1}^{1,2}$) are avaiable for all.} 
    \label{fig:architecture}
\end{figure}

Our proposed protocol is designed as a three-party architecture (Figure~\ref{fig:architecture}). It involves two spacecraft operators, designated as $\mathcal{O}_1$ and $\mathcal{O}_2$, who represent independent entities, often with conflicting interests, such as competing space agencies or commercial satellite operators. Additionally, we assume the existence of a third-party cloud server, designated as $\mathscr{C}$. This entity can be a dedicated cloud server. Their goal is to compute $\mathcal{P}_c$ between two independent spacecraft without revealing private data like position, velocity, or other sensitive parameters. To achieve this, both operators participate in an MPC protocol $\Pi = \Pi(\pi_1, \cdots, \pi_3)$ as well as the server, who assists in the protocol by carrying out encrypted operations.

We assume the most elementary threat model -- the semi-honest adversarial model -- within the context of secure MPC involving three entities. In this model, each entity holds a private input, and the goal is to jointly compute a function over their private inputs without revealing additional information beyond the final result. Under the semi-honest model, we assume that all entities correctly follow the prescribed protocol; however, adversarial entities may attempt to infer additional information from the information exchanged during protocol execution. In our specific scenario, two satellite operators and a server want to jointly compute the probability of collision $\mathcal{P}_{c}$ between their two satellites. The inputs represent private orbital quantities held by each satellite operator, as well as additional data used by the server. The proposed protocol $\Pi$, Encrypted $\mathcal{P}_{c}$, aims to ensure that the satellite operators' private data remains confidential, with only the final result $\mathcal{P}_{c}$ being revealed at the end.

\subsubsection{Brief Introduction to HE}
Our protocol relies heavily on the use of HE to facilitate secure computations. Thus, we briefly review the concept in this subsection. HE is a cryptographic primitive that allows computations to be performed directly on encrypted data without needing to decrypt it. This ensures that sensitive information remains confidential throughout the entire computation process. The key advantage of HE lies in its ability to maintain data confidentiality while enabling valid computation results. When applied within the MPC framework, the synergy between HE and MPC could enable collaborative computations that are not possible when using either method alone.

A HE scheme \cite{gentry2009fully} can be explained as a tuple $\mathcal{E} = (\mathsf{KeyGen}, \mathsf{Enc}, \mathsf{Dec}, \mathsf{Eval})$, where $\mathsf{KeyGen}$ is the key generation algorithm that produces a public-private key pair $(\mathsf{pk}, \mathsf{sk})$, $\mathsf{Enc}$ is the encryption algorithm that encrypts a message using the public key $\mathsf{pk}$, $\mathsf{Dec}$ is the decryption algorithm that decrypts the ciphertext using the secret key $\mathsf{sk}$, and $\mathsf{Eval}$ is the evaluation algorithm that allows computations to be performed on encrypted data. In the proposed architecture, two HE cryptosystems are used, both sharing the same structure, with the $i$-th HE cryptosystem defined by the tuple $\mathcal{E}_{i}$. Both cryptosystems may utilize the same algorithms but are initialized with different key pairs. The distinction between the two cryptosystems lies in the different ownership of their respective secret keys. The key generation algorithm $\mathsf{KeyGen}$ is a publicly available algorithm, which outputs a pair of keys $(\mathsf{pk}, \mathsf{sk})$, and each spacecraft owner can run the algorithm and retain its secret key $\mathsf{sk}$ while distributing their public keys to the public. We assume both HE cryptosystems to be at least leveled-homomorphic\footnote{An encryption scheme is leveled-homomorphic if it allows a limited number of encrypted multiplications. On the other hand, it it allows an unlimited number of encrypted multiplications, we call the HE scheme fully homomorphic \cite{gentry2009fully}.} encryption schemes.

Our proposed protocol should not be restricted to a particular construction of HE even though our protocol design was based on the computational capabilities available by the specific scheme such as the GSW-LWE scheme \cite{chillotti2016faster, KIM2022200}. However, since the focus of this paper is on the conceptual development of a secure MPC protocol for computing $\mathcal{P}_{c}$, from here onwards, we abstract away from a specific construction to keep the protocol presentation more general. We will use the following abstractions of algorithms involved in the protocol:
\begin{itemize}
    \item $p = \mathsf{encode}(m)$: a message $m$ encoded to plaintext $p$.    
    \item $c = \mathsf{Enc}(p, \mathsf{pk})$: a plaintext $p$ encrypted, yielding $c$.
    \item $p = \mathsf{Dec}(c, \mathsf{sk})$: a ciphertext $c$ decrypted, yielding $p$.
    \item $c_{\text{add}} = c_{1} \oplus c_{2}$: encrypted addition of two ciphertexts $c_{1}$ and $c_{2}$, resulting in another ciphertext, $c_{\text{add}} $. We assume additive homomorphism, that is, $\mathsf{Dec}(c_{\text{add}}, \mathsf{sk}) = \mathsf{Dec}(c_{1}, \mathsf{sk}) + \mathsf{Dec}(c_{2}, \mathsf{sk})$.
    \item $c_{\text{mult}} = c_{1} \otimes c_{2}$: encrypted multiplication of two ciphertexts $c_{1}$ and $c_{2}$, resulting in another ciphertext, $c_{\text{mult}} $. We assume multiplicative homomorphism, that is, $\mathsf{Dec}(c_{\text{mult}}, \mathsf{sk}) = \mathsf{Dec}(c_{1}, \mathsf{sk}) \cdot \mathsf{Dec}(c_{2}, \mathsf{sk})$.
    \item $c_{\text{sum}} = \bigoplus_{i=1,\cdots, k}c_{i}$: encrypted summation of $k$ ciphertexts by repetitive use of $\oplus$.
    \item $c_{\text{prod}} = \bigotimes_{i=1, \cdots, k}c_{i}$: encrypted product of $k$ ciphertexts by repeitive use of $\otimes$.
\end{itemize}

\begin{figure*}[h] 
    \centering    \label{fig_main}
    \includegraphics[width=2.12\columnwidth]{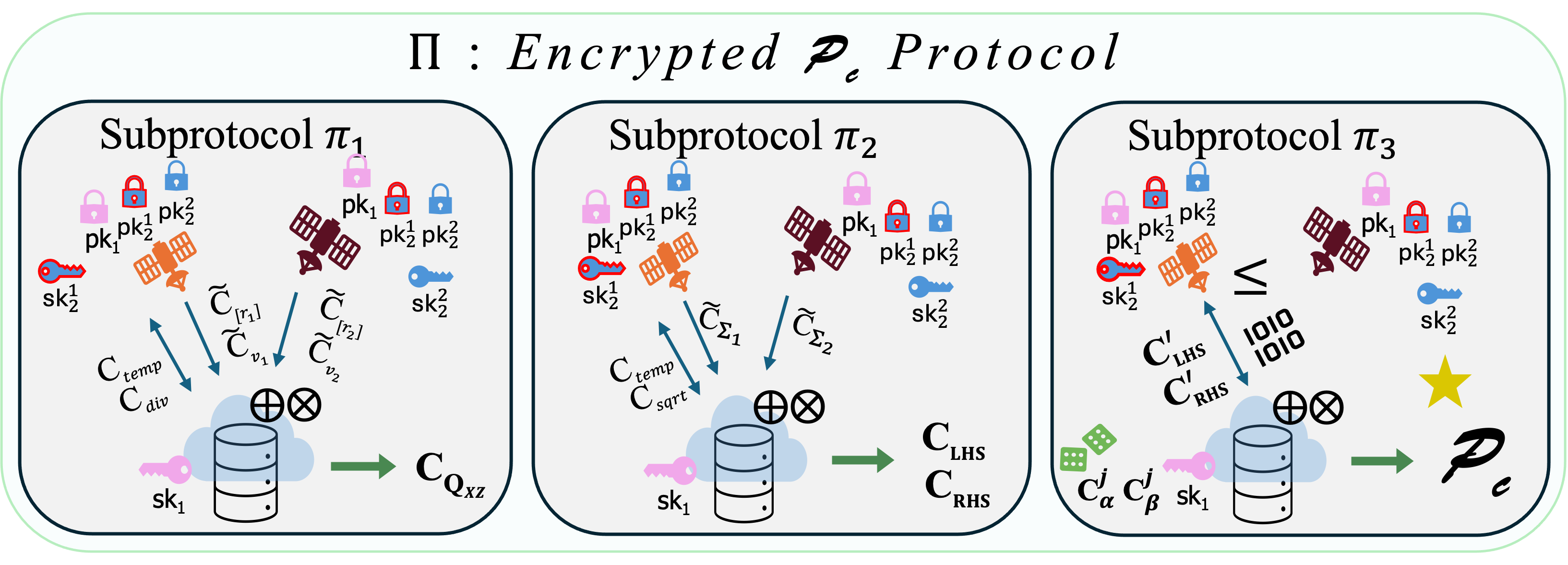}
    \caption{Pictorial illustration of $\Pi:$ Encrypted $\mathcal{P}_c$ with its subroutines $\pi_{1, 2, 3}$.}
\end{figure*}

\subsubsection{$\Pi$: Encrypted $\mathcal{P}_c$ -- and subprotocols $\pi_{1, 2, 3}$}
Let us assume for now that privacy concerns do not exist. In this safe setting, a single designated entity, e.g., a centralized server, can apply the basic Monte Carlo simulation approach to assess the collision risk between two spacecraft owned or operated by different entities during a close approach. More specifically, since both spacecraft operators can fully disclose information about their spacecraft positions and associated uncertainties, the central server gains access to the combined covariance matrix, $\Sigma_{c, \text{XZ}}$, and can accordingly generate samples $s^j = \begin{bmatrix} s_{x}^{j} & s_{z}^{j} \end{bmatrix}^{\top}$.
Subsequently, the central server has access to the value $r_0 = \|r_{rel}\| = \|r_1 - r_2\|$, along with $R = R_1 + R_2$, since the two operators also share their spacecraft's size and position data. With this information, the central server can straightforwardly evaluate the following inequality:
\begin{equation} \label{inequality}
\begin{bmatrix}
    s_{x}^{j} - r_0 \\ s_{z}^{j}
\end{bmatrix}^{\top}\begin{bmatrix}
    s_{x}^{j} - r_0 \\ s_{z}^{j}
\end{bmatrix} \leq R^2, 
\end{equation}
count how many samples satisfy this condition, and then determine $\mathcal{P}_c$.

In the real world, where spacecraft operators are reluctant to share their positions for privacy reasons, we need privacy-preserving protocols that can achieve the above outcome. Under the privacy constraints, the central server can face critical computational challenges because it has limited access to some private quantities: $r_1, r_2, R_1, R_2, \Sigma_1$, and $\Sigma_2$. 

To overcome this privacy constraint, we designed the protocol $\Pi$, which consists of a series of subprotocols $\pi_{1, \cdots, 3}$, each with a specific output:
\begin{itemize}
    \item[$\pi_{1}$] $\Rightarrow$ ciphertext of the transformation matrix $Q_{\text{XZ}}$
    \item[$\pi_{2}$] $\Rightarrow$ ciphertexts for both sides of $\eqref{inequality}$,
    \item[$\pi_{3}$] $\Rightarrow$ estimation of $\mathcal{P}_c$ via a secure comparison protocol.
\end{itemize}

Let us begin by describing the first Subprotocol $\pi_1$. The primary objective of $\pi_1$ is to enable $\mathscr{C}$ to compute $Q_{\text{XZ}}$ without knowing individual quantities $r_1, r_2, v_1,$ and $v_2$. It is important to note that these individual quantities are time-synchronized\footnote{Time synchronization is crucial since the data are encrypted, and using quantities from different time epochs could lead to incorrect results.} at the epoch when the server issues a data submission request. The subprotocol begins by collecting encrypted data to compute the relative quantities such as $r_{rel}$ and $v_{rel}$ via encrypted additions. These relative quantities can be used to compute the cross product in \eqref{rel}. Note that a cross product between $a = \begin{bmatrix}a_1 & a_2 & a_3\end{bmatrix}$ and $b = \begin{bmatrix}b_1 & b_2 & b_3\end{bmatrix}$ can be easily computed by the following matrix-vector multiplication
\begin{equation}
    a \times b := [a]^{\text{ss}}b,
\end{equation}
\noindent where \begin{equation}[a]^{\text{ss}}:= \small{\begin{bmatrix}
0 & -a_3 & a_2 \\
a_3 & 0 & -a_1 \\
-a_2 & a_1 & 0
\end{bmatrix}}\end{equation} is the skew-symmetric embedding of $a$. We also have
\begin{equation} \label{skew_symm_prop}
[a]^{\text{ss}} - [b]^{\text{ss}} = [a-b]^{\text{ss}},
\end{equation} which can be useful for casting the encrypted evaluation of the cross-product into a simple encrypted multiplication. Next, we implement a pseudo-encrypted division by the use of multiplicative masking to compute the components that make up the reference transformation matrix $Q$ in \eqref{transform}. Finally, the ciphertext of the reference transformation matrix $Q$ can be projected onto the conjunction plane, and the projection can be evaluated via plaintext-ciphertext multiplication.

\begin{algorithm}
\caption{\textbf{Subprotocol $\pi_{\arabic{algorithm}}$ -- Part 1}}\label{alg1}
\begin{algorithmic}[1]
\Require Server ($\mathscr{C}$), Operator 1 ($\mathcal{O}_1$), Operator 2 ($\mathcal{O}_2$)
\Ensure $\mathscr{C}$ obtains $\mathsf{C}_{Q_{\text{XZ}}}$, $\: i=1,2, \:j=1, \cdots, N.$
\State \textbf{Step 1: Preparation}
    \State $\mathcal{O}_{i}$ generates sets of keys $(\mathsf{pk}_{2}^{i}, \mathsf{sk}_{2}^{i})$.
    \State $\mathscr{C}$ publishes $b(j) \leftarrow \{1, 2\}$.
    \State $\mathscr{C}$ draws a random number $w$
    \State $\mathscr{C}$ prepares the multiplicative masks using  $w$ and $w^2$: 
    $$\mathsf{C}_{w^2} = \mathsf{Enc}(w^2, \mathsf{pk}_{2}^{b(j)}), \quad \mathsf{C}_{w} = \mathsf{Enc}(w, \mathsf{pk}_{2}^{b(j)}).$$
    \State $\mathscr{C}$ prepares the encoding of the projection map $\mathbb{P}_{\text{XZ}}$:
    $$\mathbb{P}_{\text{XZ}} = \begin{bmatrix}
        1 & 0 & 0 \\ 0 & 0 & 1
    \end{bmatrix}, \quad M_{\mathbb{P}} = \mathsf{encode}(\mathbb{P}_{\text{XZ}})$$
\State \textbf{Step 2: Encryption and Transfer}
    \State $\mathcal{O}_i$ performs double encryptions:
    \begin{align}
        \mathsf{\tilde{C}_{[r_i]}} &= \mathsf{Enc}(\mathsf{Enc}([(-1)^{(i-1)}r_i]^{\text{ss}}, \mathsf{pk}_{2}^{b(j)}), \mathsf{pk}_{1}) \nonumber\\
        \mathsf{\tilde{C}_{v_i}} &= \mathsf{Enc}(\mathsf{Enc}((-1)^{(i-1)}v_i, \mathsf{pk}_{2}^{b(j)}, \mathsf{pk}_{1}) \nonumber
    \end{align}
\State \textbf{Step 3: Encryption Addition}
    \State $\mathscr{C}$ decrypts the outer encryption $\mathsf{Enc}(\cdot, \mathsf{pk}_{1})$:
        \[
        \mathsf{C}_{(\cdot)} = \mathsf{Dec}(\mathsf{\tilde{C}_{(\cdot)}}, \mathsf{sk}_{1})
        \]
    \State $\mathscr{C}$ performs encrypted additions:
    \[\mathsf{C_{[r_{rel}]}} = \mathsf{C_{[r_1]}} \oplus \mathsf{C_{[r_2]}} = \mathsf{Enc}([r_{rel}]^{\text{ss}}, \mathsf{pk}_{2}^{b(j)})\]
    \[\mathsf{C_{v_{rel}}} = \mathsf{C_{v_1}} \oplus \mathsf{C_{v_2}} = \mathsf{Enc}(v_{rel}, \mathsf{pk}_{2}^{b(j)})\]
\State \textbf{Step 4: Encrypted Cross Product}
    \State $\mathscr{C}$ performs the encrypted multiplication:
    \[
    \mathsf{C}_{h} = \mathsf{C_{r_{rel}}} \otimes \mathsf{C_{v_{rel}}}
    \] 
\end{algorithmic}
\end{algorithm}

\begin{algorithm}
\caption{\textbf{Subprotocol $\pi_{1}$ -- Part 2}}\label{alg1}
\begin{algorithmic}[1]
\State \textbf{Step 5: Pseudo Encrypted Division}
    \State $\mathscr{C}$ computes the masked the inner product $\mathsf{C}_{\text{temp}}$:
        $$
        \mathsf{C}_{\text{temp}} = \mathsf{Enc}(w^{2}||v_{rel}||^2, \mathsf{pk}_{2}^{b(j)}) = \mathsf{C_{v_{rel}}} \otimes \mathsf{C_{v_{rel}}} \otimes \mathsf{C}_{w^2}
        $$
    and send it to $\mathcal{O}_{b(j)}$.
    \State $\mathcal{O}_{b(j)}$ decrypts, takes an inverse of its square root, and encrypts again, then sends it back to $\mathscr{C}$:
        \begin{align}
            \mathsf{C}_{\text{div}} &= \mathsf{Enc}\bigg(\dfrac{1}{w||v_{rel}||}, \mathsf{pk}_{2}^{b(j)}\bigg) \nonumber \\ &= \mathsf{Enc}\bigg(\big(\mathsf{Dec}(\mathsf{C}_{\text{temp}}, \mathsf{sk}_{2}^{b(j)})\big)^{-\frac{1}{2}}, \mathsf{pk}_{2}^{b(j)}\bigg) \nonumber
        \end{align}
    
    \State $\mathcal{O}_1, \mathcal{O}_2$ and $\mathscr{C}$ perform \textbf{Steps 2-3} of $\pi_{1}$ -- \textbf{Part $1$} to encrypt the skew-symmetric embedded $v_{rel}$:
        \[
        \mathsf{C}_{[v_{rel}]} = \mathsf{Enc}([v_{rel}]^{\text{ss}}, \mathsf{pk}_{2}^{b(j)})  
        \]
    \State $\mathscr{C}$ removes the mask with $\mathsf{C}_{w}$ and $\mathsf{C}_{[v_{rel}]}$:
        \[
        \mathsf{C}_{[Y]} = \mathsf{Enc}( \bigg[ \dfrac{v_{rel}}{||v_{rel}||} \bigg]^{\text{ss}}, \mathsf{pk}_{2}^{b(j)}) = \mathsf{C}_{[v_{rel}]} \otimes \mathsf{C}_{\text{div}} \otimes \mathsf{C}_{w}
        \]
        \[
        \mathsf{C}_{Y} = \mathsf{Enc}( \dfrac{v_{rel}}{||v_{rel}||}, \mathsf{pk}_{2}^{b(j)}) = \mathsf{C}_{v_{rel}} \otimes \mathsf{C}_{\text{div}} \otimes \mathsf{C}_{w}
        \]
    \State $\mathscr{C}$ redraws $w$, and prepare new masks:
    $$\mathsf{C}_{w^2} = \mathsf{Enc}(w^2, \mathsf{pk}_{2}^{b(j)}), \quad \mathsf{C}_{w} = \mathsf{Enc}(w, \mathsf{pk}_{2}^{b(j)}).$$
    \State $\mathscr{C}$ retrieves $\mathsf{C}_{h}$ from Part 1, perform the masking process as previously done:
        $$
        \mathsf{C}_{\text{temp}} = \mathsf{Enc}(w^{2}||h||^2, \mathsf{pk}_{2}^{b(j)}) = \mathsf{C_{h}} \otimes \mathsf{C_{h}} \otimes \mathsf{C}_{w^2}
        $$
    and send it to $\mathcal{O}_{b(j)}$.
    \State $\mathcal{O}_{b(j)}$ and $\mathscr{C}$ repeat Lines 3-5 (without skew-symmetric embedding parts) and $\mathscr{C}$ acquires:
        \[
        \mathsf{C}_{Z} = \mathsf{Enc}(\dfrac{h}{||h||}, \mathsf{pk}_{2}^{b(j)}) = \mathsf{C_{h}} \otimes \mathsf{C}_{\text{div}} \otimes \mathsf{C}_{w}
        \]
\State \textbf{Step 6: Encrypted Cross Product}
    \State $\mathscr{C}$ performs the encrypted multiplication: 
        \[
        \mathsf{C}_{X} = \mathsf{C}_{[Y]} \otimes \mathsf{C}_{Z}
        \]
\State \textbf{Step 7: Construction of $\mathsf{C}_{Q_{\text{XZ}}}$}
    \State $\mathscr{C}$ constructs the encrypted transformation matrix $\mathsf{C}_{Q}$ and performs the plaintext-ciphertext multiplication:
        \[
        \mathsf{C}_{Q} = \begin{bmatrix}
            \mathsf{C}_{X} & \mathsf{C}_{Y} & \mathsf{C}_{Z} 
        \end{bmatrix}^{\top}, \quad \mathsf{C}_{Q_{\text{XZ}}} = M_{\mathbb{P}}\mathsf{C_{Q}}.
        \]
\end{algorithmic}
\end{algorithm}

After the execution of Subprotocol $\pi_{1}$, $\mathscr{C}$ has the encryption of $Q_{\text{XZ}}$. On the other hand, operators $\mathcal{O}_1$ and $\mathcal{O}_2$ have their respective covariance matrices\footnote{Both operators are assumed to hold positive definite covariances and agree on the same matrix factorization method during the setup phase.}, $\Sigma_1$ and $\Sigma_2$. The main objective of Subprotocol $\pi_2$ is to help $\mathscr{C}$ obtain the ciphertexts that represent the operands of \eqref{inequality}.

Recall that the combined covariance is the sum of individual covariances when two distributions are independent and the linearity of transformations \eqref{cov_transform}-\eqref{cov_proj} on the Gaussian random variable. Let us first outline the strategy for the second Subprotocol $\pi_{2}$ without considering any encrypted computations. Each operator $\mathcal{O}_i$ holds the matrix square root of its respective covariance. Independent standard normal samples $z_{i}^{j}\sim\mathcal{N}(0, I)$ can be first transformed by these square roots:
\begin{equation}\label{first_transf}
    \tilde{s}_{i}^{j}  = \Sigma_{i}^{\frac{1}{2}}z_{i}^{j}, \quad \text{for } i=1, 2, \: j=1,\cdots, N,
\end{equation}
such that $\tilde{s}_{i}^{j} \sim \mathcal{N}(0, \Sigma_{i})$, followed by the rotating and projecting onto the conjunction plane via $Q_{\text{XZ}}$ such that
\begin{equation} \label{third_transf}
    s_{i}^{j} = Q_{\text{XZ}}\tilde{s}_{i}^{j}, \quad \text{for } i=1, 2, \: j=1,\cdots, N,
\end{equation}
with each sample $s_{i}^{j} \sim \mathcal{N}(0, \Sigma_{i, \text{XZ}} := Q_{\text{XZ}}\Sigma_{i}Q_{\text{XZ}}^{\top})$. Furthermore, they can be summed up so that
\begin{equation}\label{second_transf}
    s^{j}  = {s}_{1}^{j} + {s}_{2}^{j}, \quad \text{for } j=1,\cdots, N.
\end{equation}
The resulting samples follow ${s}^{j} \sim \mathcal{N}(0, \Sigma_{c, \text{XZ}})$ since  
\begin{equation}\label{cov_transform_XZ}
    \Sigma_{c, \text{XZ}} = \underbrace{Q_{\text{XZ}} \Sigma_{1} Q_{\text{XZ}}^{\top}}_{=\Sigma_{1, \text{XZ}}} +\underbrace{Q_{\text{XZ}} \Sigma_{2} Q_{\text{XZ}}^{\top}}_{=\Sigma_{2, \text{XZ}}} = Q_{\text{XZ}}\Sigma_{C}Q_{\text{XZ}}^{\top}.
\end{equation}
Subprotocol $\pi_{2}$ is designed to enable the transformations \eqref{first_transf}-\eqref{cov_transform_XZ} over HE, leaving $\mathscr{C}$ with encrypted samples $\mathsf{C}_{s}^{j}$.

\begin{algorithm}
\caption{\textbf{Subprotocol $\pi_{2}$}}\label{alg2}
\begin{algorithmic}[1]
\Require Server ($\mathscr{C}$), Operator 1 ($\mathcal{O}_1$), Operator 2 ($\mathcal{O}_2$)
\Ensure $\mathscr{C}$ obtains $\mathsf{C}_{\text{LHS}}$, $\mathsf{C}_{\text{RHS}}$, $i=1,2$, $j=1, \cdots, N$.
\State \textbf{Step 1: Preparation}
    \State $\mathcal{O}_{i}$ computes $\Sigma_{i}^{\frac{1}{2}}$, and generates keys $(\mathsf{pk}_{2}^{i}, \mathsf{sk}_{2}^{i})$.
    \State $\mathscr{C}$ posts $b(j) \leftarrow \{1, 2\}$, samples $z_{i}^{j} \sim \mathcal{N}(0, I)$, encodes $\mathsf{M}_{-1}:=\mathsf{encode}(\begin{bmatrix}
        -1 & 0
    \end{bmatrix}^{\top})$.
\State \textbf{Step 2: Covariance Transformation}
    \State $\mathcal{O}_{i}$ performs double encryptions, transmits to $\mathscr{C}$: 
    \begin{align}
        \mathsf{\tilde{C}}_{\Sigma_{i}} &= \mathsf{Enc}(\mathsf{Enc}(\Sigma_{i}^{\frac{1}{2}}, \mathsf{pk}_{2}^{b(j)}), \mathsf{pk}_1)\nonumber \\
        \mathsf{\tilde{C}}_{R_{i}} &= \mathsf{Enc}(\mathsf{Enc}(R_{i}, \mathsf{pk}_{2}^{b(j)}), \mathsf{pk}_{1}) \nonumber
    \end{align}
    \State $\mathscr{C}$ decrypts the outer encryption $\mathsf{Enc}(\cdot, \mathsf{pk}_{1})$:
    \[
    \mathsf{C}_{(\cdot)} = \mathsf{Dec}(\mathsf{\tilde{C}_{(\cdot)}}, \mathsf{sk}_{1})
    \]
    and obtains $\mathsf{C}_{\Sigma_{i}} = \mathsf{Enc}(\Sigma_{i}^{\frac{1}{2}}, \mathsf{pk}_{2}^{b(j)})$, and $\mathsf{C}_{R_{i}}$.
    \State $\mathscr{C}$ performs encrypted multiplication and acquires:
    $$\mathsf{C}_{\Sigma_{i, \text{XZ}}} = \mathsf{C}_{Q_{\text{XZ}}} \otimes\mathsf{C}_{\Sigma_{i}} = \mathsf{Enc}(Q_{\text{XZ}}\Sigma_{i}^{\frac{1}{2}}, \mathsf{pk}_{2}^{b(j)})$$
\State \textbf{Step 3: Calculate the miss-distance}
    \State $\mathscr{C}$ draws $w$ and masks, and send to $\mathcal{O}_{b(j)}$:
    $$\mathsf{C}_{\text{temp}} = \mathsf{C}_{w^2} \otimes \mathsf{C_{r_{rel}}} \otimes \mathsf{C_{r_{rel}}} = \mathsf{Enc}(w^{2}||r_0||^{2}, \mathsf{pk}_{2}^{b(j)}).$$
    \State $\mathcal{O}_{b(j)}$ decrypts it, takes a square root, encrypts:
    \begin{align}
        \mathsf{C}_{\text{sqrt}} &= \mathsf{Enc}((\mathsf{Dec}(\mathsf{C}_{\text{temp}}, \mathsf{sk}_{2}^{b(j)}))^{\frac{1}{2}},\mathsf{pk}_{2}^{b(j)})\nonumber
    \end{align}
    and return it to $\mathscr{C}$.
    \State $\mathscr{C}$ computes the encrypted miss-distance component
    \[
    \mathsf{C}_{r_0} = \mathsf{M}_{-1} \mathsf{C}_{\text{sqrt}} = \mathsf{Enc}(-w\begin{bmatrix}
        r_0 \\ 0
    \end{bmatrix},\mathsf{pk}_{2}^{b(j)}),
    \]
\State \textbf{Step 4: Sampling and Encryption}
    \State $\mathscr{C}$ encrypts the masked samples and transform
    $$\mathsf{C}_{s}^{j} = \bigoplus_{i=1,2}\bigg(\mathsf{C}_{\Sigma_{i, \text{XZ}}} \otimes \mathsf{Enc}(w{z}_{i}^{j}, \mathsf{pk}_{2}^{b(j)})\bigg).$$
\State \textbf{Step 5: Obtain the LHS and the RHS}
    \State $\mathscr{C}$ performs the encrypted addition:
    \[
    \mathsf{C}_{\text{LHS}}^{j} = \mathsf{C}_{s}^{j} \oplus \mathsf{C}_{r_{0}}, \quad \mathsf{C}_{R} = \mathsf{C}_{R_1}\oplus\mathsf{C}_{R_2}.
    \]
    \State $\mathscr{C}$ performss the encrypted multiplication (masking):
    \[
    \mathsf{C}_{\text{RHS}}^{j} = \mathsf{Enc}(w^{2}, \mathsf{pk}_{2}^{b(j)})\otimes\mathsf{C}_{R}.
    \]
\end{algorithmic}
\end{algorithm}

Additionally, $\mathscr{C}$ must also perform the encrypted subtraction of $r_0$ from the encrypted samples it already holds. To this end, we again employ a delegated square root operation after protecting $||r_0||^2 = {r_{rel}}^{\top} r_{rel}$ with multiplicative masking, followed by a series of encrypted multiplications and additions to construct $\mathsf{C}_{\text{LHS}}^{j}$, the scaled LHS of the inequality \eqref{inequality}. The ciphertext of the scaled RHS, $\mathsf{C}_{\text{RHS}}^{j} = \mathsf{Enc}(w^{2}R^2, \mathsf{pk}_{2}^{b(j)})$, can be computed with only a few encrypted computations. After completing Subprotocol $\pi_{2}$, the only task remaining is to evaluate the following relationship without decrypting either ciphertext
\begin{equation}\label{cipher_comparison}
\mathsf{C}_{\text{LHS}} \leq \mathsf{C}_{\text{RHS}},
\end{equation}
and successfully evaluating \eqref{cipher_comparison} enables an accurate estimation of $\mathcal{P}_{c}$ as its decryption represents the scaled \eqref{inequality}:
\begin{equation} \label{scaled_inequality}
w^{2}\begin{bmatrix}
    s_{x}^{j} - r_0 \\ s_{z}^{j}
\end{bmatrix}^{\top}\begin{bmatrix}
    s_{x}^{j} - r_0 \\ s_{z}^{j}
\end{bmatrix} \leq w^{2}R^2.
\end{equation}

Unfortunately, a direct comparison is not easy due to the shuffling effect introduced by encryption. The comparison of two ciphertexts is a variant (due to operands being encrypted) of the broader problem, the secure integer comparison, often called Yao's Millionaires' problem \cite{yao1982protocols}. Potential solutions include the DGK protocol \cite{damgaard2007efficient}, which uses bit-wise encryption, and an approximate comparison method \cite{cheon2019numerical} that involves homomorphic evaluation of the sigmoid function. When an efficient bootstrapping algorithm is available, \cite[Algorithm 2]{bourse2020improved} is a promising protocol, which evaluates the sign function on the difference of comparison operands. Thus, we may assume \cite[Algorithm 2]{bourse2020improved} as a direct subroutine available for the Subprotocol $\pi_{3}$ if efficient bootstrapping is available at the server.

\begin{algorithm}
\caption{\textbf{Subprotocol $\pi_{3}$}}
\label{alg4}
\begin{algorithmic}[1]
\Require Server ($\mathscr{C}$), Operator 1 ($\mathcal{O}_1$), Operator 2 ($\mathcal{O}_2$)
\Ensure $\mathcal{P}_{c}$
\State \textbf{Step 1: Preparation}
    \State $\mathscr{C}$ draws $b(j) \leftarrow \{1, 2\}$.
    \State $\mathscr{C}$ samples pairs of random numbers $\{\alpha^{j}, \beta^{j}\}$.
    \State $\mathscr{C}$ encrypts each pair: 
    \[
    \mathsf{C}_{\alpha}^{j} = \mathsf{Enc}(\alpha^{j}, \mathsf{pk}_{2}^{b(j)}), \quad \mathsf{C}_{\beta}^{j} = \mathsf{Enc}(\beta^{j}, \mathsf{pk}_{2}^{b(j)})
    \]
\State \textbf{Step 2: Masking the Comparison}
    \State $\mathscr{C}$ performs the encrypted masking:
    \[
    \mathsf{C}_{\text{LHS}}' = \mathsf{C}_{\alpha}^{j} \otimes \mathsf{C}_{\text{LHS}} \oplus \mathsf{C}_{\beta}^{j}, \quad \mathsf{C}_{\text{RHS}}' = \mathsf{C}_{\alpha}^{j} \otimes \mathsf{C}_{\text{RHS}} \oplus \mathsf{C}_{\beta}^{j}
    \]
\State \textbf{Step 3: Delegation to Operator}
    \State $\mathscr{C}$ sends the masked ciphertexts $\mathsf{C}_{\text{LHS}}'$ and $\mathsf{C}_{\text{RHS}}'$ to the selected operator $\mathcal{O}_{b(j)}$.
    \State $\mathcal{O}_{b(j)}$ decrypts masked ciphertexts $\mathsf{C}_{\text{LHS}}'$ and $\mathsf{C}_{\text{RHS}}'$:
    \begin{align*}
    \mathsf{Dec}(\mathsf{C}_{\text{LHS}}', \mathsf{sk}_{2}^{b(j)}) &= \alpha^{j} \mathsf{Dec}(\mathsf{C}_{\text{LHS}}, \mathsf{sk}_2) + \beta^{j} \\
    \mathsf{Dec}(\mathsf{C}_{\text{RHS}}', \mathsf{sk}_{2}^{b(j)}) &= \alpha^{j} \mathsf{Dec}(\mathsf{C}_{\text{RHS}}, \mathsf{sk}_2) + \beta^{j}
    \end{align*}
\State \textbf{Step 4: Reporting the Result}
    \State $\mathcal{O}_{b(j)}$ returns the comparison result to $\mathscr{C}$.
\State \textbf{Step 5: Inference by the server}
    \State $\mathscr{C}$ infers the result of the original comparison.
\State \textbf{Step 6: Compute the collision probability}
    \State $\mathscr{C}$ computes $\mathcal{P}_{c}$ by counting the results from \textbf{Step 5}.
\end{algorithmic}
\end{algorithm}

However, when efficient bootstrapping is unavailable, we can explore a masking-based alternative solution, which is proposed in the Subprotocol $\pi_{3}$. Let us recall that the operators possess the secret key $\mathsf{sk}_{2}^{i=1,2}$. Therefore, a straightforward delegation of the comparison of two ciphertexts $\mathsf{C}_{\text{LHS}}$ and $\mathsf{C}_{\text{RHS}}$ to one of the operators is not secure because they can be decrypted. To prevent the operator from decrypting these two ciphertexts, the server can draw pairs of two random numbers $\alpha^{j}, \beta^{j} \neq 0$, and encrypt them under the public key $\mathsf{pk}_{2}^{b(j)}$, producing $\mathsf{C}_{\alpha}^{j} = \mathsf{Enc}(\alpha^{j}, \mathsf{pk}_{2}^{b(j)})$ and $\mathsf{C}_{\beta}^{j} = \mathsf{Enc}(\beta^{j}, \mathsf{pk}_{2}^{b(j)})$. The server performs the masking on the original inequality
\begin{equation} \label{masked_comparison}
\mathsf{C}_{\alpha}^{j} \otimes \mathsf{C}_{\text{LHS}} \oplus \mathsf{C}_{\beta}^{j} \leq \mathsf{C}_{\alpha}^{j} \otimes \mathsf{C}_{\text{RHS}} \oplus \mathsf{C}_{\beta}^{j},
\end{equation}
which can then be delegated to the operator $i=b(j)$ for decryption, and evaluation. This way, the server collects the boolean values for $N$ comparisons from the operator and finally computes the probability of collision by simple counting of comparison results.

\section{Discussion and Future Work}\label{discussion}
The methodology proposed in this paper---integrating HE, and Monte Carlo estimation within the MPC framework---has significant potential for applications beyond secure satellite conjunction analysis. Any domain requiring secure, privacy-preserving computations among multiple parties, especially where sensitive data must be collaboratively analyzed without exposure, can benefit from the methods presented in this paper. Potential applications span diverse sectors such as finance, healthcare, and energy. Furthermore, each application is likely to present unique constraints, driving further innovations in the development of the secure protocol to meet specific privacy and performance needs.

In designing our proposed protocol, the existence of a politically neutral cloud server was assumed to simulate a non-colluding model. However, we recognize that such a trusted server may not always be practical, and this assumption should be carefully examined, and alternatives should be explored. In future work, we will investigate potential improvements to the protocol towards the elimination of such servers, perhaps by the use of other cryptographic primitives or decentralized frameworks.

Furthermore, we will conduct a detailed security analysis of the proposed protocol. Also, we plan to revisit the comparison protocol for potential improvements in its complexity. We can also examine the computational and communication efficiency of the proposed protocol in comparison with existing protocols in the literature.

\subsection{Conclusion} \label{conclusion}
In this work, we presented Encrypted $\mathcal{P}_c$, a novel secure protocol for computing the probability of collision between two satellites while preserving the privacy of sensitive orbital data. By leveraging creative uses of HE within the three-party MPC architecture proposed, we designed a secure protocol that enables stakeholders to collaborate on computing $\mathcal{P}_{c}$ without revealing their proprietary or strategically sensitive information.

Our proposed method utilizes a Monte Carlo estimation algorithm for the secure computation protocol, providing an alternative to existing secure computation protocols that revolve around the evaluation of integral using MPC. We presented the design of the main protocol $\Pi$ through a series of subprotocols $\pi_1$, $\pi_2$, and $\pi_3$. The proposed solution demonstrates the potential for applying advanced cryptographic techniques to enable secure satellite conjunction analysis, and furthermore, secure SSA, which is becoming increasingly important as the space environment becomes increasingly congested.

\newpage



\bibliography{space.bib}{}
\bibliographystyle{IEEEtran}

\end{document}